\documentclass[prc,12pt,english]{revtex4}
\usepackage{babel}
\usepackage{amsmath} 
\usepackage{amssymb} 
\usepackage{graphicx} 
\usepackage{graphics} 
\usepackage{epsfig} 
\usepackage{slashed}
\usepackage{multirow}

\usepackage{pdfpages} 

\newcommand\ba{\begin{eqnarray}}
\newcommand\ea{\end{eqnarray}}
 \newcommand\alb{\begin{align}}
\newcommand\ale{\end{align}} 
\newcommand\be{\begin{equation}}
\newcommand\ee{\end{equation}}

\usepackage[T2A]{fontenc}
\usepackage[utf8]{inputenc}
\definecolor{buccia}{RGB}{3,125,3}
\definecolor{polpa}{RGB}{217,60,46}
\definecolor{seme}{RGB}{101,67,33}

\begin{document}

\title{At the heart of baryonic matter the \textbf{\textit{\textcolor{buccia}{А}\textcolor{polpa}{р}\textcolor{seme}{б}\textcolor{polpa}{у}\textcolor{buccia}{з}}} model favored by recent hadron form factor data 
}

\author{E.~Tomasi-Gustafsson} \email{egle.tomasi@cea.fr}
\affiliation{\it DPhN, IRFU, CEA, Universit\'e Paris-Saclay, 91191
Gif-sur-Yvette Cedex, France}

\author{Simone~Pacetti } \email{simone.pacetti@unipg.it}
\affiliation{\it Dipartimento di Fisica e Geologia, and INFN Sezione di
Perugia, 06123 Perugia, Italy }

\begin{abstract} 
Data on electromagnetic form factors of proton,  neutron,  and $\Lambda$  from annihilation and scattering reactions are collected and interpreted in the frame of a generalized picture of the internal structure of baryons which holds in space-like and time-like regions.  It is  shown that these data give an insight of the space structure of the baryon for distances one hundred times smaller than the baryon size and,  in the time-like  region,  a vision of the time evolution of the hadronic matter for times up to $10^{-25}$~s, that is two orders of magnitude shorter than the time taken by the light to cross the volume of a proton.  In the proposed interpretation,  the electric form factor of the proton in the space-like region can not cross zero, but vanishes or stays very small, as an extrapolation of the data seems to show. In the time-like region specific structures appearing in the data give evidence of the dominance of the quark-diquark structure in a specific range of time of the evolution of the system. 
\end{abstract}
\date{\today}
\maketitle 

\section{Introduction}
\label{Section:introduction}
Electromagnetic form factors (FFs) contain essential information on the internal structure of hadrons and their study constitutes a basic field of the intermediate energy physics domain, experimentally as well as theoretically.  In a parity  and time invariant theory a particle of spin $S=1/2$ is characterized by two FFs, electric $(G_E)$ and magnetic ($G_M$),  that are functions of one variable only (for a review see Ref. \cite{Pacetti:2015iqa}). The information of FFs is accessible through elementary scattering and annihilation reactions, assuming one photon exchange. The FF values are extracted from the data (differential cross section and polarization observables) corrected by all experimental corrections, after taking into account radiative corrections. 

The elastic scattering of polarized electrons on an unpolarized proton target with the measurement of the polarization of the recoil proton $\vec e^-+p\to e^-+\vec p$,  provides the information on FFs in the space-like region of transferred momenta ($q^2 < 0$, being $q^2=-Q^2$ the four momentum of the virtual photon).  This method, suggested by A.I. Akhiezer and M.P. Rekalo at the end of 1960's \cite{Akhiezer:1968ek,Akhiezer:1974em} was applied only recently, due to the development of high duty cycle electron machines, large acceptance detectors and hadron polarimeters in the GeV region.  The results on the ratio of the electric to magnetic FF, as expected, were  very precise,  due to the sensitivity of the method to the small electric contribution. Surprisingly,  they showed that this ratio is not constant, but decreases almost linearly with $Q^2$, eventually crossing zero around $Q^2\simeq 9$ GeV$^2$.
 
In the time-like region,  accessible in annihilation reactions,   regular oscillations were highlighted in Ref.  \cite{Bianconi:2015owa} considering the precise data collected by BABAR \cite{Lees:2013uta,Lees:2013xe} on the proton generalized FF,  later on confirmed by the BESIII Collaboration in Ref.  \cite{BESIII:2019hdp,BESIII:2021rqk}.  The BESIII Collaboration published also the first individual determination of the moduli of the electric and magnetic proton FF in the time-like region \cite{BESIII:2019hdp},  and  unique data on the neutron FF,   analyzed similarly to the proton, in terms of oscillations with similar characteristics,  but shifted by a phase  \cite{BESIII:2021tbq}.

The purpose of this work is to analyze and interprete the recent data obtained on one side in the time-like region on neutron, proton and hyperons most recently by the BESIII collaboration \cite{BESIII:2019hdp,BESIII:2021rqk}, but also by BABAR \cite{Lees:2013uta,Lees:2013xe} and by CMD \cite{CMD-3:2018kql},  and on the other side in the space-like region mostly by the GEp Collaboration \cite{Puckett:2017flj}, in the framework of a model suggested ten years ago in Ref. \cite{Kuraev:2011vq}.  The basic assumption of the model is that the spacial center of the hadron is electrically neutral due to the strong gluonic field.   This assumption has two principal effects: to induce a screening that traslates into a suppression of the electric FF with respect to the magnetic one,  and to favour the development of a diquark configuration during the evolution of the system from the quark creation to the hadron formation.  These features should be present in both space-like and time-like regions, although the physical meaning of FFs differs in these domains.  

In the space-like region, FFs have a clear interpretation in non relativistic approximation,  where they are the Fourier transforms of the electric charge and magnetic spatial density distributions.  This holds also in the Breit system, where the energy of the virtual photon is zero, and its four-momentum reduces to a three-momentum, as in the non relativistic case, $q=(0,q_x,q_y,q_z)$.  Therefore in the space-like region, FFs contain information related to the spatial densities in the proton at the scale defined by the four-momentum of the virtual photon. 

In the time-like region,  the privileged system is the Center of Mass System (CMS), where the three momentum is zero, i.e.,  $q=(q_0,  0,0,0)$.  Only the time component of the transferred momentum plays a role. A time-like FF can not bring any spatial information, as, in the process $e^+e^-\to \bar p p $, it describes the time evolution of the charge created at the annihilation point until the formation of the detected hadron. Eventually, this scale can be associated to the distance of the centers of the forming hadrons.   

In order to formalize these concepts,  a generalized definition has been introduced and developped in Ref.  \cite{Kuraev:2011vq}. Form factors are functions of $q^2$ only, therefore it is possible to define a relativistic invariant in the following way
\be
F(q^2)=\int_{\cal D} d^4x\  e^{iq_{\mu}x^{\mu}}\rho(x), \ \ 
q_{\mu}x^{\mu}=q_0t-\vec q\cdot\vec x,
\label{eq:eq5b}
\ee
where $\rho(x)=\rho(t,\vec x) $ can be understood as the space-like distribution of the electric charge in a space-time volume ${\cal D}$.

In the scattering channel, $ep\to ep$, and in the Breit frame, we recover the usual definition of FFs
\be
F(q^2)=F(- \vec q ^2), 
\ee
where zero energy transfer is implied.  

In the annihilation channel  and in CMS we have
\be
F(q^2)=\int_{{\cal D}_t} dt \  e^{i\sqrt{q^2} t }\int_{{\cal D}_x} d^3 \vec x \ \rho(x)=
\int_{{\cal D}_t}  dt \  e^{i\sqrt{q^2} t } {\cal Q}(t)\,,
\ee
where ${\cal Q}(t)$ describes the time evolution of the charge distribution in the temporal subset ${\cal D}$, i.e., 
$\forall t \in {\cal D}_t $ obtained after integration of the space-time distribution $\rho(x)$ over  the spatial subdomain 
${\cal D}_x$, having the decomposition  ${\cal D}={\cal D}_t \cup {\cal D}_x$.
 
This means that experimentally, we have access to the projections of the generalized function on the space and on the time axis, in the Breit system and in  CMS, respectively.
  
In the next Section we recall the main features of the model of Ref.  \cite{Kuraev:2011vq},  in Section III the space-like data are presented as a function of the internal spatial dimension as seen by the virtual photon,  
while the time-scale is illustrated in Section IV for the time-like data of nucleons and hyperons.  In Section V a remarkable correlation among these FFs is discussed.  

\section{Description of the scattering and annihilation processes}
\label{Section:description}
The nucleon description in terms of constituent quarks or vector dominance models assumes that the three colored valence quarks are surrounded  by a neutral sea of quark-antiquark  pairs and gluons.  The model of Ref.  \cite{Kuraev:2011vq}  gives a different picture based on studies of the structure of QCD vacuum and gluon condensate \cite{Vainshtein:1982zc}. 

The center volume of the nucleon is assumed to be chromo-electrically neutral,  due to the strong gluonic field that creates a gluon condensate, with a randomly oriented chromomagnetic field \cite{Vainshtein:1982zc}. At very small distances the gluon field as well as the chromoelectric field increases, inducing a screening effect that acts on the electric FF, leaving the magnetic distribution unchanged,  similarly to the Coulomb field in a plasma.  The magnetic distribution is expected to follow a $Q^2$ dipole dependence,  according to the scaling rules of QCD,  while it can be shown that the electric distribution is suppressed by an extra factor of $1/Q^2$. 

In the region of strong chromo-electromagnetic field, due to stochastic averaging, the color quantum number does not play any role.  Therefore, due to the Pauli principle, quarks of the same flavor,  $uu$ for proton and $dd$ for  neutron,  leave the central region,  and one of them is attracted by the remaining quark, $d$ in the proton  and $u$ in the neutron, forming a diquark.  As the system expands and cools down, the strength of the gluon field decreases and the color degree of freedom  is restored.  This step is driven by the balance of the electric attraction force and the stochastic force of the gluon field.  It is predicted to occur at a distance of  0.2-0.3 fm.   At larger distances the gluon energy transforms into 'dressing' the quarks,  that convert into constituent quarks.

The hadron formation in $e^+e^- $ annihilation can also be described through three main steps in terms of evolution in time.  In order to create the hadron-antihadron pair,  the energy at the $e^+e^-$ annihilation, concentrated in a small volume, should be at least equal or larger than the threshold energy, $E_{Th}=2M_h$ ($M_h$ being the hadron and antihadron mass).  Then,  $q\bar q$ pairs are created by the vacuum fluctuations, with the same probability independently on flavor.  However,  due to the uncertainty principle, the time associated with the $q\bar q$ pair depends on their mass and hence on the flavor, the heavier the  $q\bar q$ pair, the shorter the formation time.  This affects the probability to create a hadron-antihadron pair, which requires  for the $p\bar p$ final channel,  that two pairs  $u\bar u$ and one pair $d\bar d$ are created in a  space-time volume of dimensions $ [\hbar/(2M_h)]^3 \simeq$ (0.1 fm)$^3$.  Below the physical threshold one expects that a system, constituted by at least three bare quark-antiquark pairs, is formed.  This system,  with the quantum numbers of the photon, can be considered point-like and colorless, due to  the screening of the strong chromoelectromagnetic field.  Similarly to the space-like picture,  the Pauli principle applies to the two identical quarks,  one of them  is attracted by the remaining quark, forming a diquark.  The system  expands and cools down,  the quarks absorb gluons and transform into constituent quarks with mass and magnetic moment.  The last step is the formation of the hadron-antihadron pair moving apart,  as a results of the competition between the available kinetic energy,  $T=\sqrt{q^2}-E_{Th}$ and the confinement energy,  $k_s/2 R_{pp} $, where 
$k_s\simeq 1$ GeV/fm is the confinement elasticity constant and 
 $ R_{pp}$ is the distance between the centers of the forming hadron and antihadron.  When the velocity is very small,  a bound state can be formed with dimensions up to hundreds of fm.  

In Ref. \cite{Kuraev:2011vq} the comparison with the experimental data was limited, due to the  few measurements  especially in the time-like region.  Recently an important amount of data on proton and neutron FFs in the  time-like region has been made available by the BESIII Collaboration.  In next Section we compare the dynamical evolution of the baryonic system,  as predicted by the model,  with the present world data set.

\section{The space structure of  the proton}
\label{Section:protonSL}
In the space-like region, polarization measurements provide information on the electric to magnetic FF ratio.  Instead than the usual variable $Q^2$,  we report the data from the JLab-GEp Collaboration \cite{Puckett:2017flj} as a function of $r$,  the scale length associated to the wavelength $\lambda$ of the virtual photon with squared four-momentum $Q^2$:
\be 
r\ \mbox{[fm]}=\lambda = \hbar c/\sqrt{Q^2}= 0.197\  \mbox{[GeV fm]}/ \sqrt{Q^2} \mbox{[GeV]}, 
\label{eq:eqrfm}
\ee
where small values of $r$ correspond to large four-momenta.

The internal distances covered by the kinematics where data exist extend to very small nucleon sizes,  about  two orders of magnitude smaller than the nucleon dimension.  At the other extreme, very small  values of $Q^2$ (large $r$ values) will prevent the virtual photon to resolve the proton structure,   and the electric and magnetic FFs at $Q^2=0$ have to coincide with the electric charge (one, in units of $e$ charge) and the magnetic moment $\mu_p$,  respectively.
\begin{figure} [h]
\begin{center}
\includegraphics[width=17 cm]{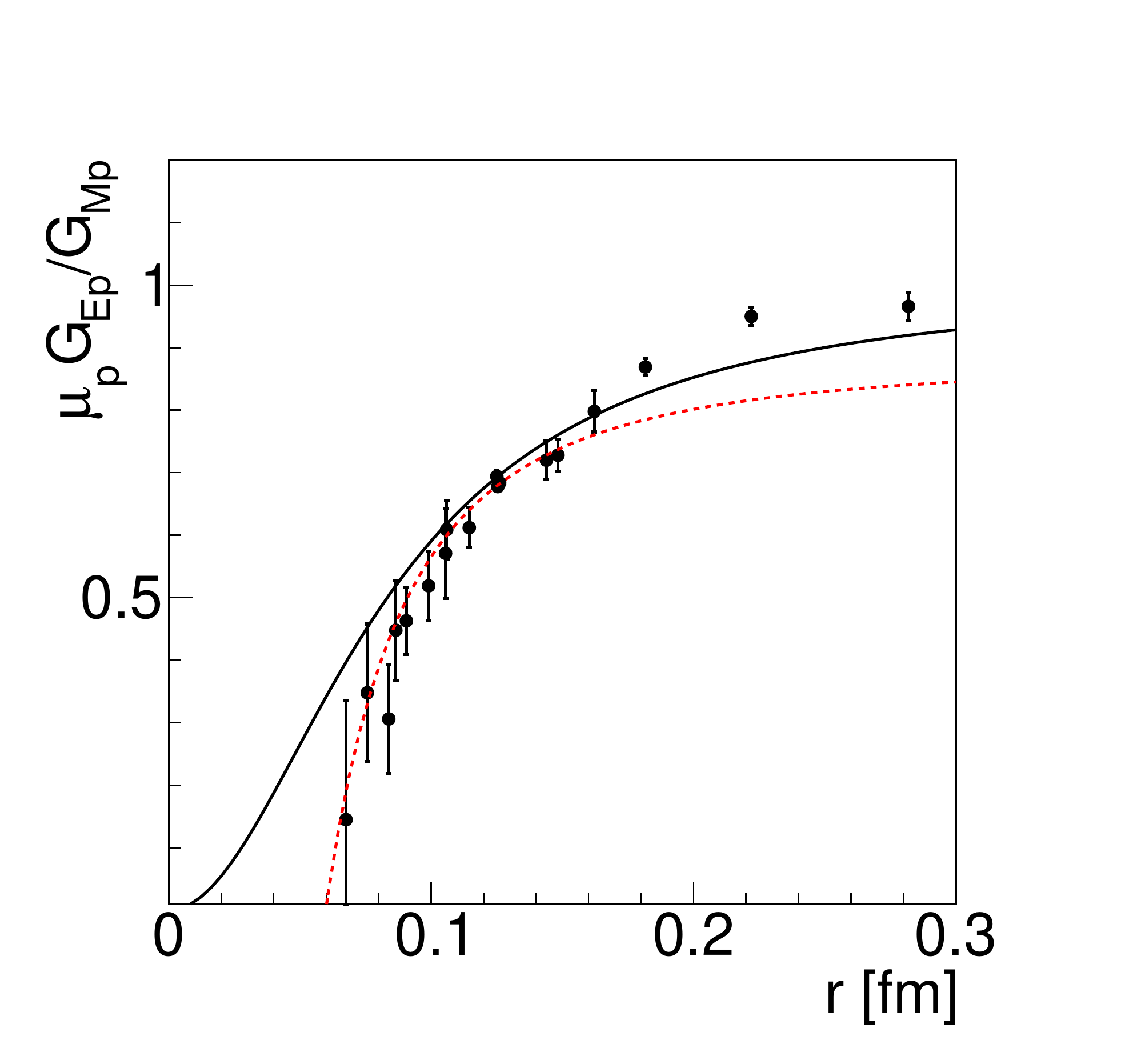} 
\caption{Proton FF ratio as a function of the internal distance seen by the virtual photon. The points are frrom Ref. \protect\cite{Puckett:2017flj} and Refs. therein. The dashed red line is a $Q^2$-linear fit of the highest $Q^2$ points, the solid black line is the monopole fit, Eq. (\protect\ref{eq:eqrSLfm}).} 
\label{Fig:rSLfm} 
\end{center} 
\end{figure}

The FF ratio ${ \cal R}=\mu_p G_{Ep}/G_{Mp}$, shown in 
Fig. \ref{Fig:rSLfm} as a function of $r$, can be parametrized by a straight line or the monopole form of mass $m_r$:
\be 
{ \cal R} =\mu_p \displaystyle\frac{G_{Ep}}{G_{Mp}}=
 \left (1+ \displaystyle\frac{ Q^2}{m_r^2} \right )^{-1}.
\label{eq:eqrSLfm}
\ee
The proton FF ratio decreases regularly,  showing definitely a suppression of $G_E$ compared to $G_M$.  It can be reasonably reproduced by 
Eq. (\ref{eq:eqrSLfm}) with $m_r^2=5.6 $  GeV$^2$ and approaches to zero, in the limit of the errors,  for $r\le 0.06$ fm. Such a distance corresponds to the largest value of $Q^2$ measured by the GEp Collaboration.  An extended program of measurements up to $Q^2$=15 GeV$^2$ is planned at Jefferson Lab, following the energy upgrade \cite{PR12-07-109}, with the main aim to investigate if the ratio will cross zero and eventually become negative.  The smooth $Q^2$-decreasing  behavior of ${\cal R}$ agrees with the model of Ref. \cite{Kuraev:2011vq}.  Moreover,  ${\cal R}$  approaching to zero at large $Q^2$ can be interpreted as due to the vanishing electric FF for internal distances approaching the  screening region.  A linear extrapolation of the high-$Q^2$ points in the $Q^2$ variable (red dashed line) allows us to define the size of the this region as about 0.06 fm.  At larger $Q^2$ values the model predicts that the ratio will stay very small.

In the light of the structures observed in the time-like region,  as well as of the inhomogeneity in the electromagnetic  density originated by the different configurations participating in the hadron-anti-hadron formation,  one may expect to observe irregularities instead of a smooth behavior of the ratio.  One reason can be that these structures have been associated to the interference among phenomena occurring at different scales or to rescattering effects related to the imaginary parts of the amplitudes.  In this case, they should be  suppressed in the space-like region, where FFs are real, appearing preferentially  in the time-like region, where FFs are complex, with non-vanishing imaginary parts due to unitarity. Another reason is that these structures may be cancelled in the FF ratio while manifest in the individual FFs.   

The Rosenbluth separation (unpolarized $ep$ elastic scattering cross section measurements at fixed $Q^2$ at different angles) allows to extract separately $G_E$ and $G_M$ but only at small $Q^2$.  At large $Q^2$ the electric contribution to the cross section is compatible with and hence hidden in the experimental error, making  doubtful the extraction of the individual FFs.  Precise polarization measurements allow to extract only the FF ratio.  But one can derive the electric FF from the ratio, assuming that the magnetic FF is well determined from the Rosenbluth measurements, as the magnetic contribution to the unpolarized cross section is dominant.  The magnetic FF has been measured up to $Q^2$=30 GeV$^2$ and it shows indeed some deviation from a dipole,  with a dip around $Q^2$= 0.2 GeV$^2$ and a bump around $Q^2$=3 GeV$^2$. 

\begin{figure} [h]
\begin{center}
\includegraphics[width=17cm]{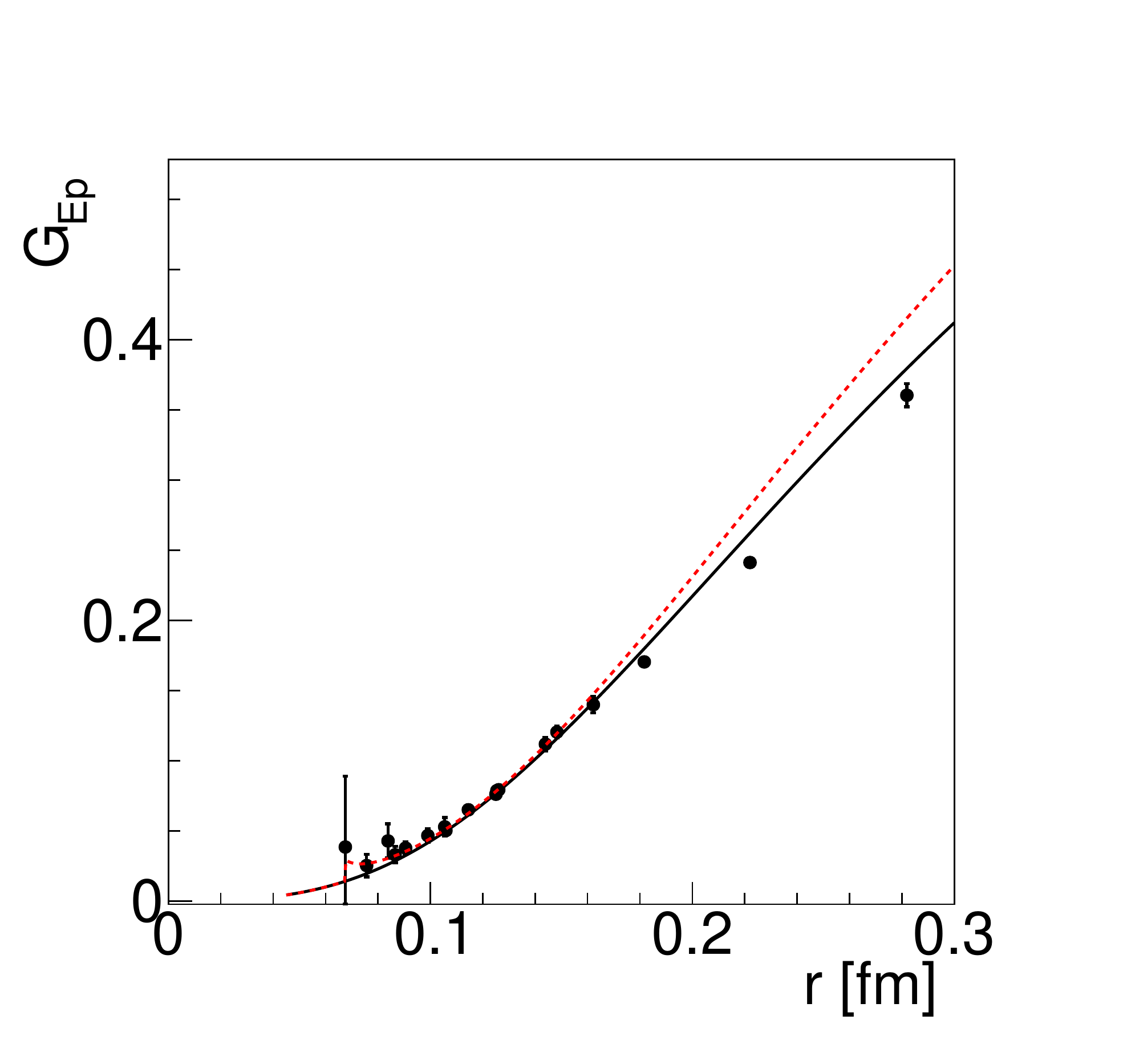} 
\caption{Electric proton FF as a function of the internal distance $r$  seen by the virtual photon. Notations as in Fig. \ref{Fig:rSLfm}.} 
\label{Fig:GEpfm} 
\end{center} 
\end{figure}

Calculating $G_{Ep}$ from the ratio ${\cal R}$,  with the help of the fit from Ref.  \cite{Brash:2001qq}, that gives a reliable description of the $G_{Mp}$ data,  one finds a smooth behavior for the electric FF,  with no evident structures as shown in Fig.  \ref{Fig:GEpfm}. 

\begin{figure} [h]
\begin{center}
\includegraphics[width=17cm]{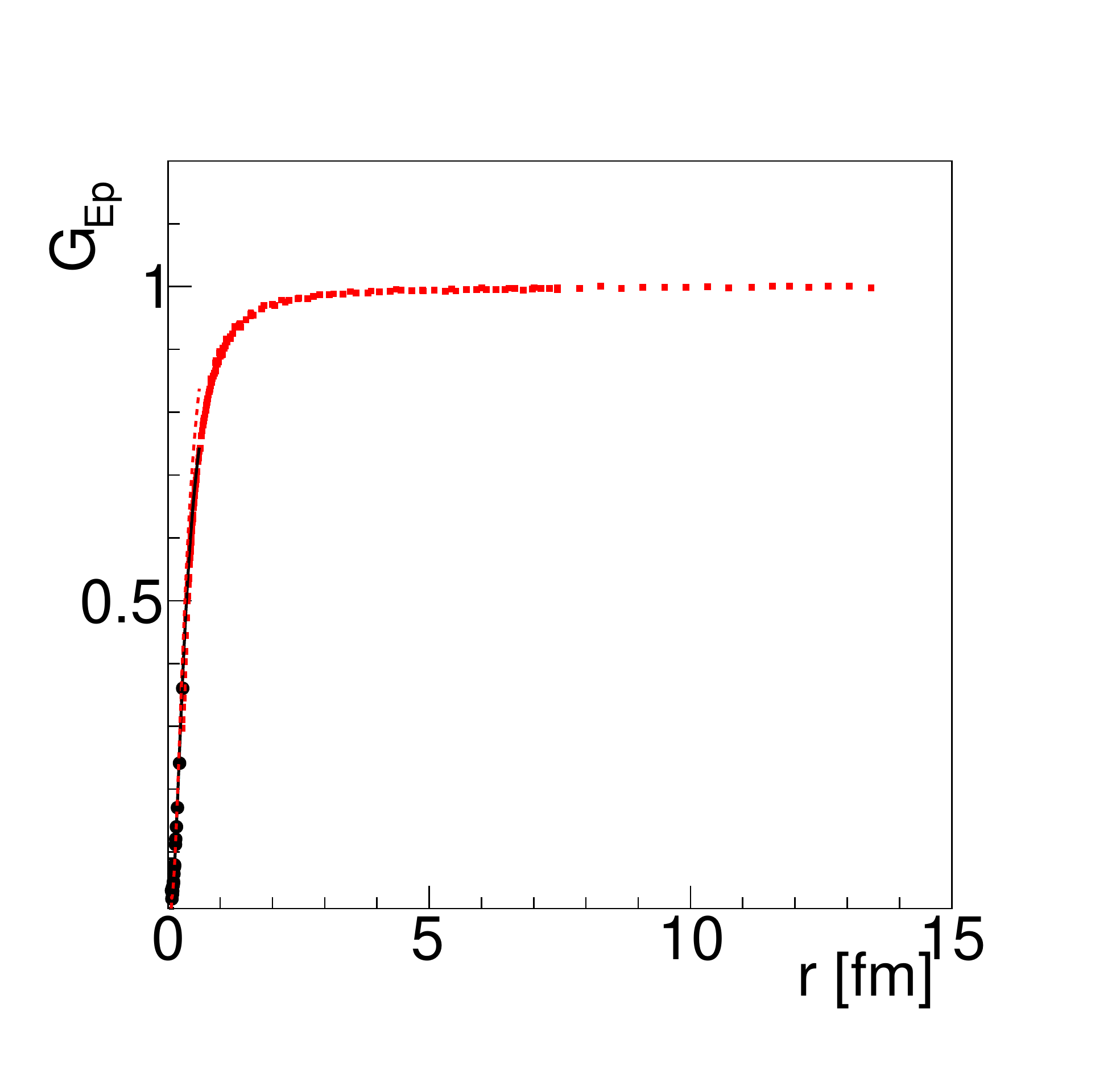} 
\caption{Electric FF of the proton as a function of the internal distance in fm, as seen by the virtual photon. The red points are from Ref. \cite{A1:2013fsc}, other notations as  in Fig. \protect\ref{Fig:GEpfm}. } 
\label{Fig:pradSLfm} 
\end{center} 
\end{figure}

The recent $G_{Ep}$ data, collected at very small $Q^2$ for determining the proton radius, can also be plotted as a function of $r$.  In Fig.  \ref{Fig:pradSLfm}  one can see that the resolving power of a photon of such a small four-momentum is very large,  up to 15 fm. Therefore one may wonder how such a photon can give a meaningful measurement of the proton dimension that is  $ ~20$  times smaller and how it can 'see' the proton size at the few permille level selecting a well defined value in the range $0.84-0.87$ fm, what is necessary to solve the so-called 'proton puzzle'.  This argument corroborates the finding and the quantitative discussion of Refs.  \cite{Pacetti:2018wwk,Pacetti:2021fji}.

Following this discussion, one can predict that  the measurements of proton FFs at large $Q^2$ values will not bring additional information on the proton structure, confirming a very small or vanishing electric FF.  The efforts to perform measurements at   $Q^2\to 0$ with higher precision is also a nonsense,  as the smaller is $Q^2$ the less the photon will see precisely the proton dimension.

\subsection{The space structure of  the neutron}
\label{Section:neutronSL}

\begin{figure} [h]
\begin{center}
\includegraphics[width=17cm]{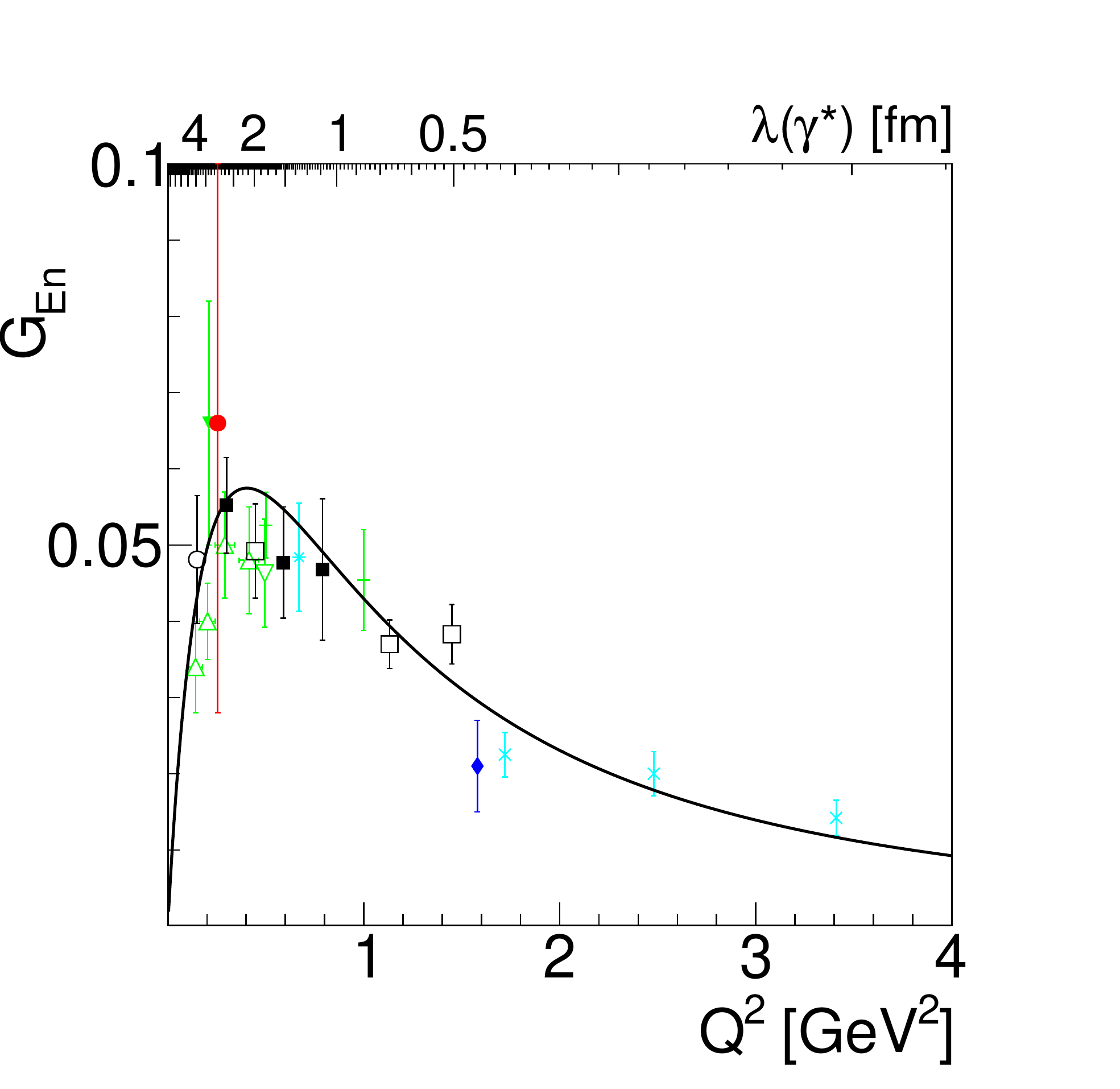} 
\caption{Electric form factor of the neutron as a function of the momentum transfer square from Refs. 
\cite{Schlimme:2013eoz} (solid blue diamond), 
\cite{Riordan:2010id} (cyan cross), 
\cite{Bermuth:2003qh} (cyan asterisk), 
\cite{Geis:2008aa} (green triangle), 
\cite{Warren:2003ma} (green cross), 
\cite{Zhu:2001md} (open green triangledown), 
\cite{Passchier:1999cj}  (solid green  triangledown), 
\cite{Plaster:2005cx} (open black square), 
\cite{Glazier:2004ny} (solid black square), \cite{Herberg:1999ud} (open solid circle), 
\cite{Eden:1994ji} (red circle).  The line is the parametrization from Ref. \cite{Kelly:2004hm}. The upper scale shows the corresponding wavelength of the virtual photon.} 
\label{Fig:GENfm} 
\end{center} 
\end{figure}

The values of the neutron electric FF in the space-like region are small with respect to unity, 
the static value (the neutron electric charge) being zero at $Q^2=0$.  It has usually been considered uniformly equal to zero,  but since that  precise data have been obtained, also using the Akhiezer-Rekalo polarization method \cite{Akhiezer:1968ek, Akhiezer:1974em}, a more complicated picture appeared.  The data on the electric neutron FF (Fig. \ref{Fig:GENfm}),  although less precise than for proton,  and extending to a shorter range,  show an increase at small $Q^2$, eventually a plateau between 0.2 and 0.8 fm, and then a decrease  to zero  at large $Q^2$ (small $r$).  The upper scale in Fig. \ref{Fig:GENfm} shows  the  corresponding wavelength of the virtual photon, i.e.,  the internal distance  that is explored. At large $Q^2$, by construction, the parametrization \cite{Kelly:2004hm} goes as $Q^{-2}$. The data show the tendency to reach a zero well above $Q^2$=4 GeV$^2$, i.e.,  for internal distances $r \ll 0.2$ fm. 

\begin{figure} [h]
\begin{center}
\includegraphics[width=17cm]{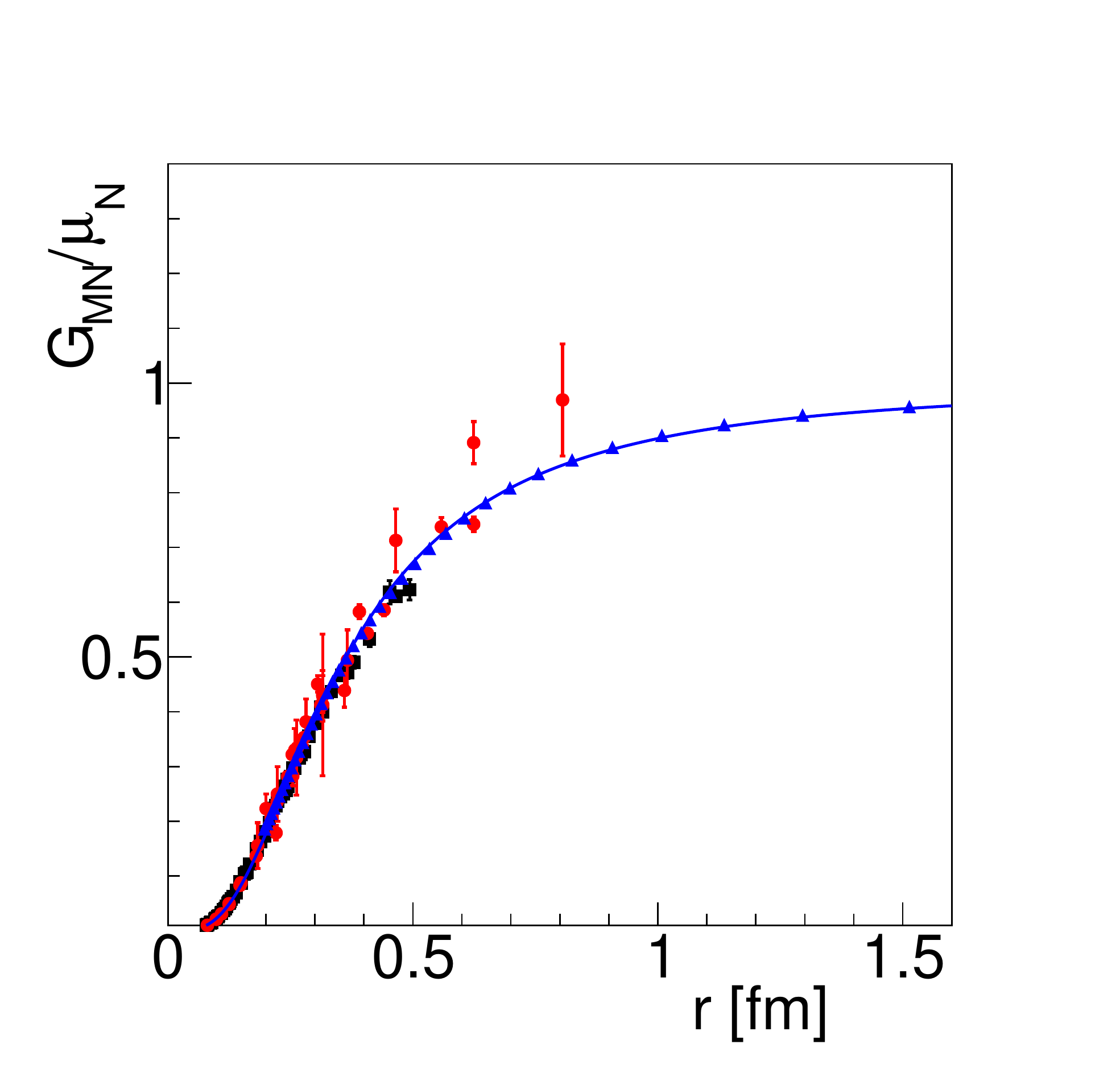} 
\caption{Magnetic FF of the proton (black squares) and of the neutron (red circles) normalized to the corrispondent magnetic moment, as a function of the internal distance $r$ in fm. The blue triangles are the $G_{Mp}$ data extracted from Ref.  \cite{A1:2013fsc} (spline option). The blue line is the dipole function.} 
\label{Fig:GMNfm} 
\end{center} 
\end{figure}

The magnetic  FFs, normalized to the corresponding magnetic moments, are  similar for neutron and proton,  essentially following the standard  dipole behavior 
\be
G_D=(1+Q^2  /0.71\ \mbox{GeV}^2 )^{-2}.
\label{eq:dipole}
\ee
In Fig. \ref{Fig:GMNfm} the magnetic FFs (normalized to the corresponding magnetic momenta) are shown as functions of $r$ for neutron (black squares) and proton (red circles). The data are selected from  the compilation of Ref. \cite{Pacetti:2015iqa}. At small $Q^2$ (large $r$) the FFs values approach unity.  The behavior at the smallest values of $Q^2$ is driven by the very precise experiments that had the aim to measure the proton radius  \cite{A1:2013fsc}. 

\section{The time structure of  protons,  neutrons and hyperons}
\label{Section:time}

Let us focus on the process $e^+e^-\to p\bar p$.  As recalled in Section  \ref{Section:description}, FFs  in the annihilation region carry information on the time evolution of the spatial distribution of the charge which is created at the annihilation point. The charge is  carried by the bare quarks that  evolve to dressed quarks and eventually diquarks,  till  the hadron-antihadron pair formation.  It may be interesting to investigate which time scale is related to these steps.   According to Ref.  \cite{Kuraev:2011vq}, the different stages of the hadron formation are related to the time of evolution of the system  and to the distance $D$  between the forming hadron and antihadron.  

\begin{figure} [h]
\begin{center}
\includegraphics[width=17cm]{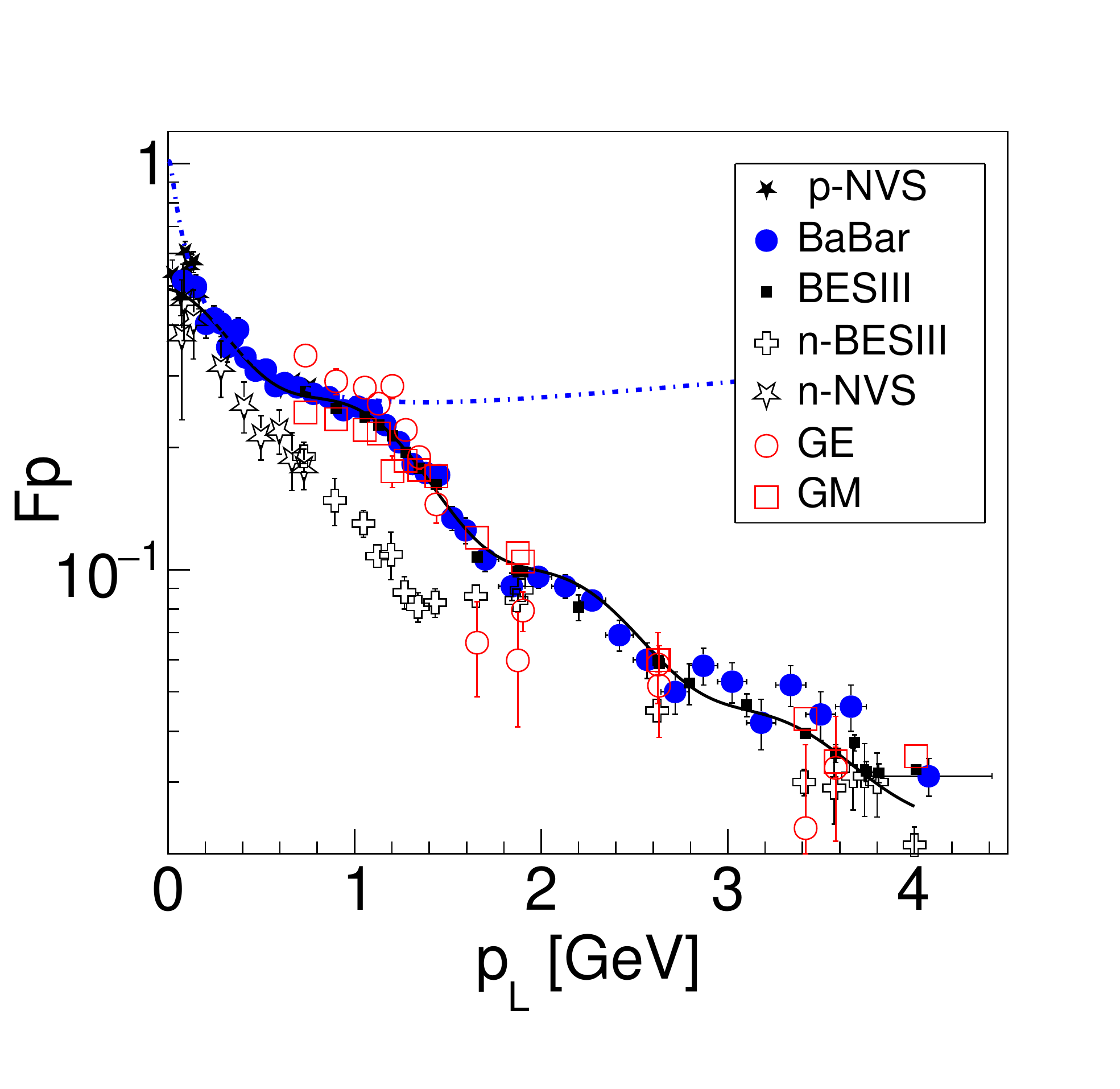} 
\caption{Time-like effective FF of the nucleon as a function of the relative momentum $p_L$:  generalized (open  black crosses),  electric (open red circles) and magnetic FF of the proton (open squares) \protect\cite{BESIII:2019hdp}.   The black solid line is the six parameter fit from Ref.    \protect\cite{Tomasi-Gustafsson:2020vae}.  The blue dashed line is the expected generalized  to a constant cross section  $\sigma=0.87$ mb,  corrresponding to  $|G_E|=|G_M|=1$.  The neutron FF is shown as open black symbols, stars from Ref.   \protect\cite{CMD-3:2018kql} and crosses from Ref.  \cite{BESIII:2021tbq}.} 
\label{Fig:allFp} 
\end{center} 
\end{figure}

The cross section of the process $e^+ e^-\to p\bar p$ has been measured by the CMD-3 Collaboration at  Novosibirsk  \cite{Akhmetshin:2015ifg, CMD-3:2018kql},  by the BaBar Collaboration at SLAC  \cite{Lees:2013xe, Lees:2013uta} and by the BESIII Collaboration at Beijing in several works, using initial state radiation \cite{BESIII:2015axk, BESIII:2021rqk} and beam scan method \cite{BESIII:2019hdp}.  The data show indeed a region where the cross section is  compatible with a structureless proton, see Fig. \ref{Fig:allFp}.  In Ref.  \cite{Tomasi-Gustafsson:2020vae} it was  found that not only the effective FF, that is a combination of the moduli of the electric and magnetic FFs  but also their ratio  shows marked oscillations,  that have to be mostly attributed to the electric FF.   

In Fig.  \ref{Fig:allFp} the effective nucleon FF is plotted as a function of the relative momentum of the produced nucleons in the laboratory system $p_L$.  Near threshold the proton and neutron FFs are comparable,  as well as above $p_L=2$ GeV.  
Regular oscillations for the proton FF, when plotted as a function of this variable, are well reproduced by the six parameter fit from Ref.  \protect\cite{Tomasi-Gustafsson:2020vae}.
The neutron FF is smaller than the proton FF in the region $ 0.4  \le p_L \le 1.4$ GeV,  it reaches a plateau for $ 1.4  < p_L \le 2$  GeV where it is about constant and then reaches the proton FF values up to large $Q^2$.  

The moduli of the electric $G_E$ and magnetic $G_M$ proton FFs are also reported in Fig. \ref{Fig:allFp}.  It appears clearly that $|G_M|$ has a smooth behavior, whereas $|G_E|$  seems to follow the behavior of the effective neutron FF, with a steep decrease and a plateau in the region $ 1.6\le p_L\le 2.6$ GeV.
At large energies  all FFs converge towards very small values.   

The time-like FF has been precisely measured by the BESIII Collaboration not only for  neutrons \cite{BESIII:2021tbq},  but also for $\Lambda$ in Ref. \cite{BESIII:2017hyw} overlapping with the data from BABAR  \cite{Aubert:2007uf},   and at higher energies  in Ref.  \cite{BESIII:2021ccp}. 

\begin{figure} [h]
\begin{center}
\includegraphics[width=17cm]{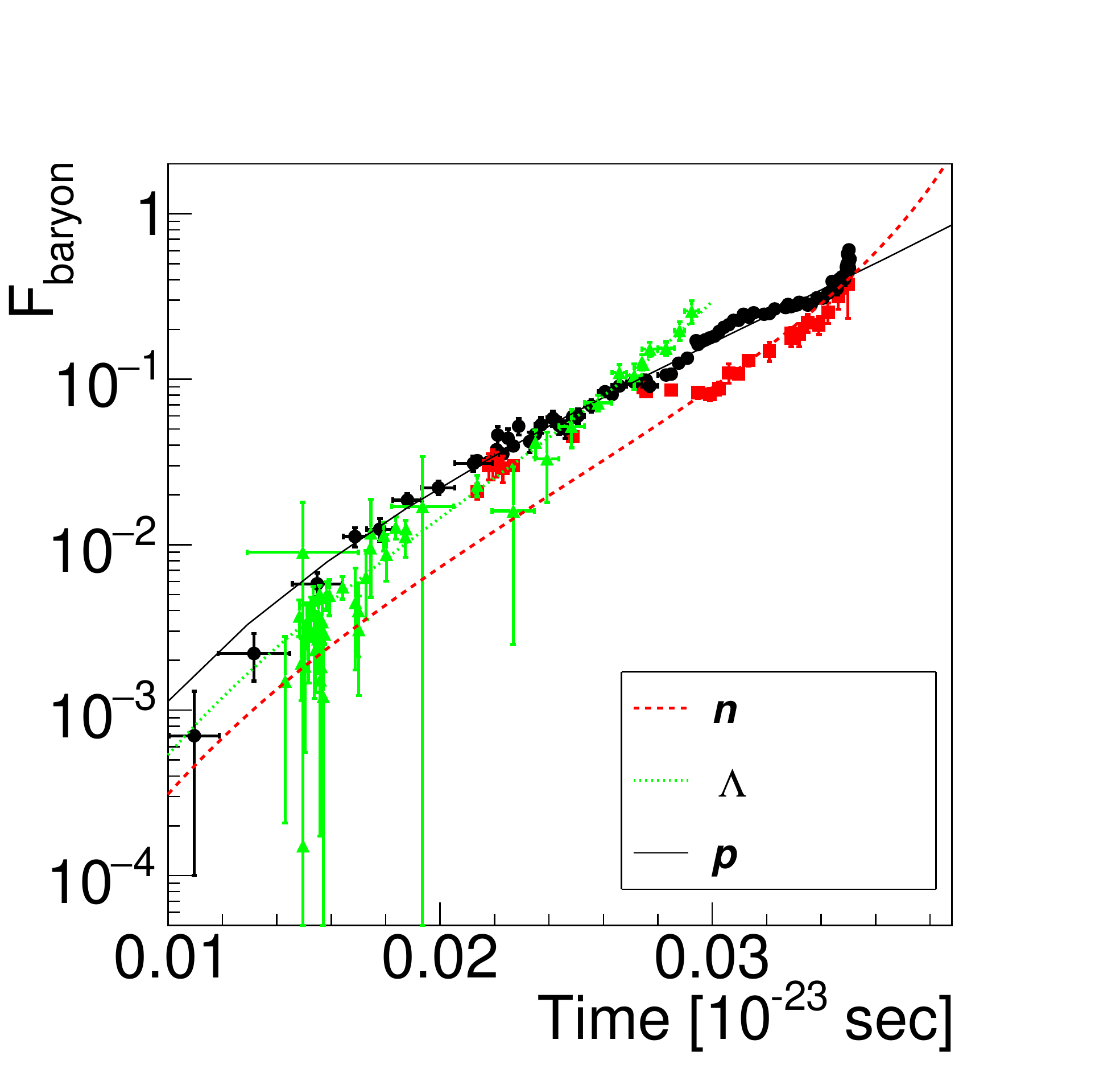} 
\caption{Time scale of FFs  as a function of the time scale for neutron \cite{BESIII:2021tbq,CMD-3:2018kql} (red squares), proton \cite{Lees:2013xe, Lees:2013uta,BESIII:2015axk} (black circles) and $\Lambda$  \protect\cite{BESIII:2017hyw,Aubert:2007uf, BESIII:2021ccp} (green triangles). } 
\label{Fig:tNPL} 
\end{center} 
\end{figure}

The transferred momentum square can be related to the time evolution from the annihilation point.  In the time-like region, and in CMS, the four momentum reduces to its energy component. Then the  energy scale can be converted in  time scale
\be 
t\ \mbox{[sec]}=\displaystyle\frac{\hbar c}{q_0}= 
\displaystyle\frac{0.0658\cdot 10^{-23} \mbox{[GeV s]}}{q_0\mbox{ [GeV]}}.
\label{eq:eqtsec}
\ee
In Fig. \ref{Fig:tNPL}  we can see that the time scale corresponding to the data is in the range $(1-3)\cdot 10^{-2}$ in units of $10^{-23}$~s,  that is the time that it takes for the light to travel through the proton.
Three main trends can be  seen: a steep decreasing near threshold (the threshold corresponds to the largest time),  a plateau that is more evident for the neutron and at comparable time for the proton,  and a dipole (or tripole) behavior at large $q^2$ (small times). The $\Lambda$ baryon follows a similar trend, the threshold and the plateau occurring at shorter times. This can be attributed to the larger mass of the $\Lambda$ and, at the quark level, to the need to create a strange quark-antiquark pair.  
 
The hadron and the antihadron move apart when the kinetic energy $T=q_0 -2M_N $  exceeds the confinement energy, $(k_s/2) R_{pp} $, where $ k_s=1$ GeV/fm  is the strength of the color force attraction. Note that in the threshold region, the dimension of the system can reach hundreds of fm.

\section{Correlation of FFs for  neutron, proton and $\Lambda$}

\begin{figure} [h]
\begin{center}
\includegraphics[width=17cm]{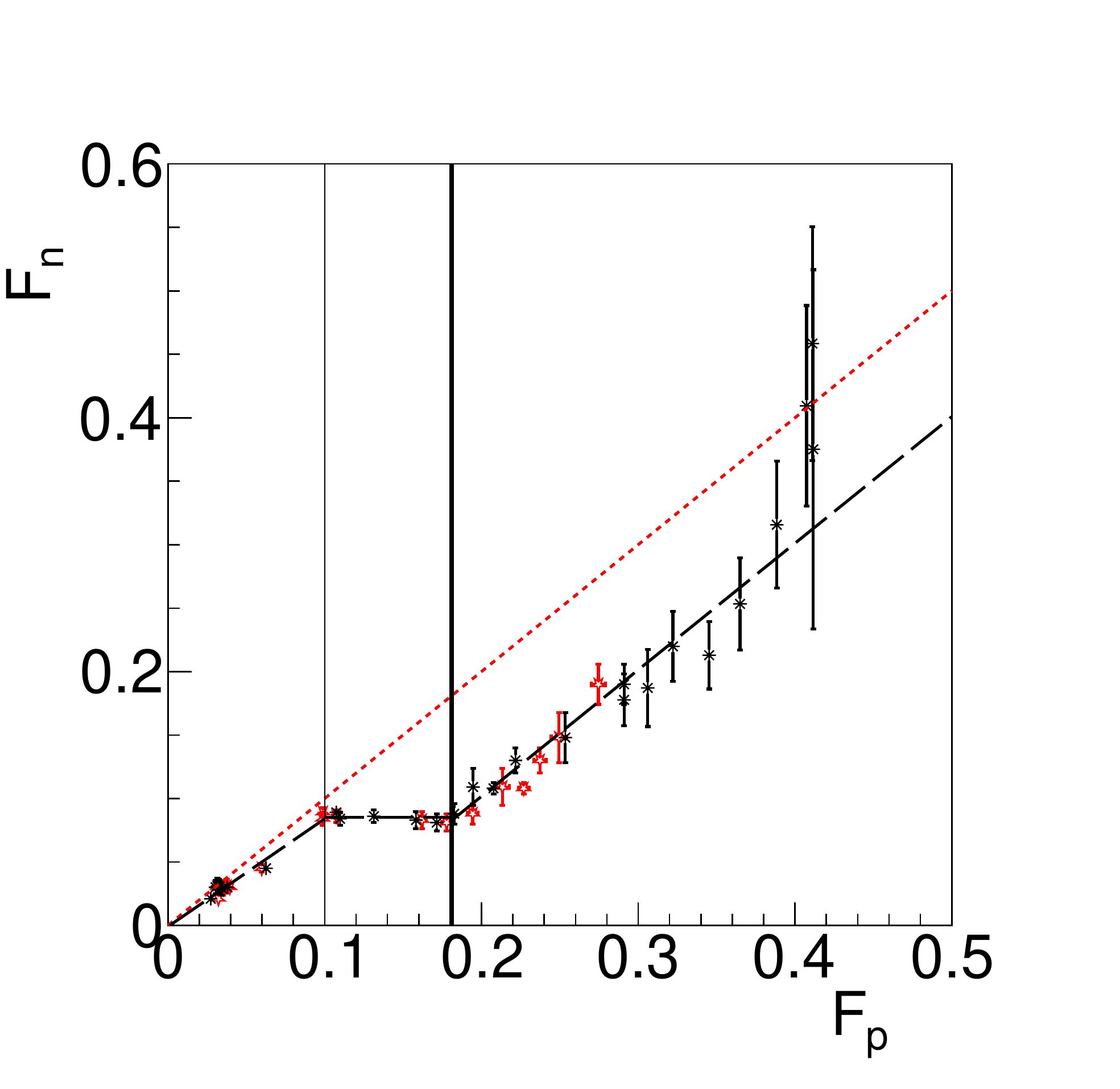} 
\caption{Neutron-proton effective FF  correlation at  corresponding $p_L$ (red open stars). When the data are not present at the same $p_L$, the six-parameter fit from Ref. \cite{Tomasi-Gustafsson:2020vae} is used for the proton. The red dashed line shows $F_n=F_p$, the black long dashed line is drawn to drive the eyes. The thin and thick black solid lines delimit the region of discontinuity (see text). } 
\label{Fig:FpFn} 
\end{center} 
\end{figure}

The fact that the three regions, corresponding to three regimes in the evolution of the baryonic system,  are common to different baryons  imply some correlation among FFs.  To make evident such correlation, in Fig.  \ref{Fig:FpFn} the effective neutron FF is plotted in the ordinate and the proton FF measured at the same $p_L$ in the abscissa (red triangles). The proton FF has also been calculated from the six-parameter proton data fit of Ref. \cite{Tomasi-Gustafsson:2020vae} (black asterisks), especially useful when data are not available at the same $p_L$. The long dashed line is drawn to guide the eyes. The dashed red line corresponds to $F_n=F_p$. 

Three regimes and two regions where the proton and neutron FF
are strongly correlated with  two breaking points indicated by the thin and thick vertical lines, are evident. The first one corresponds to $(F_p,F_n)\simeq (0.1,0.085)$, occurring around $p_L= 1.3$ GeV,  $(q^2=4.7$ GeV$^2$) and the second one at $(F_p,F_n)= (0.18,0.085)$, occurring around $p_L=1.9$ GeV, ($q^2=5.7$ GeV$^2$).
Note that the threshold corresponds to the right top of the figure and the large $q^2$ region to the points gathered near the origin.

\begin{figure}
\begin{center}
\includegraphics[width=17cm]{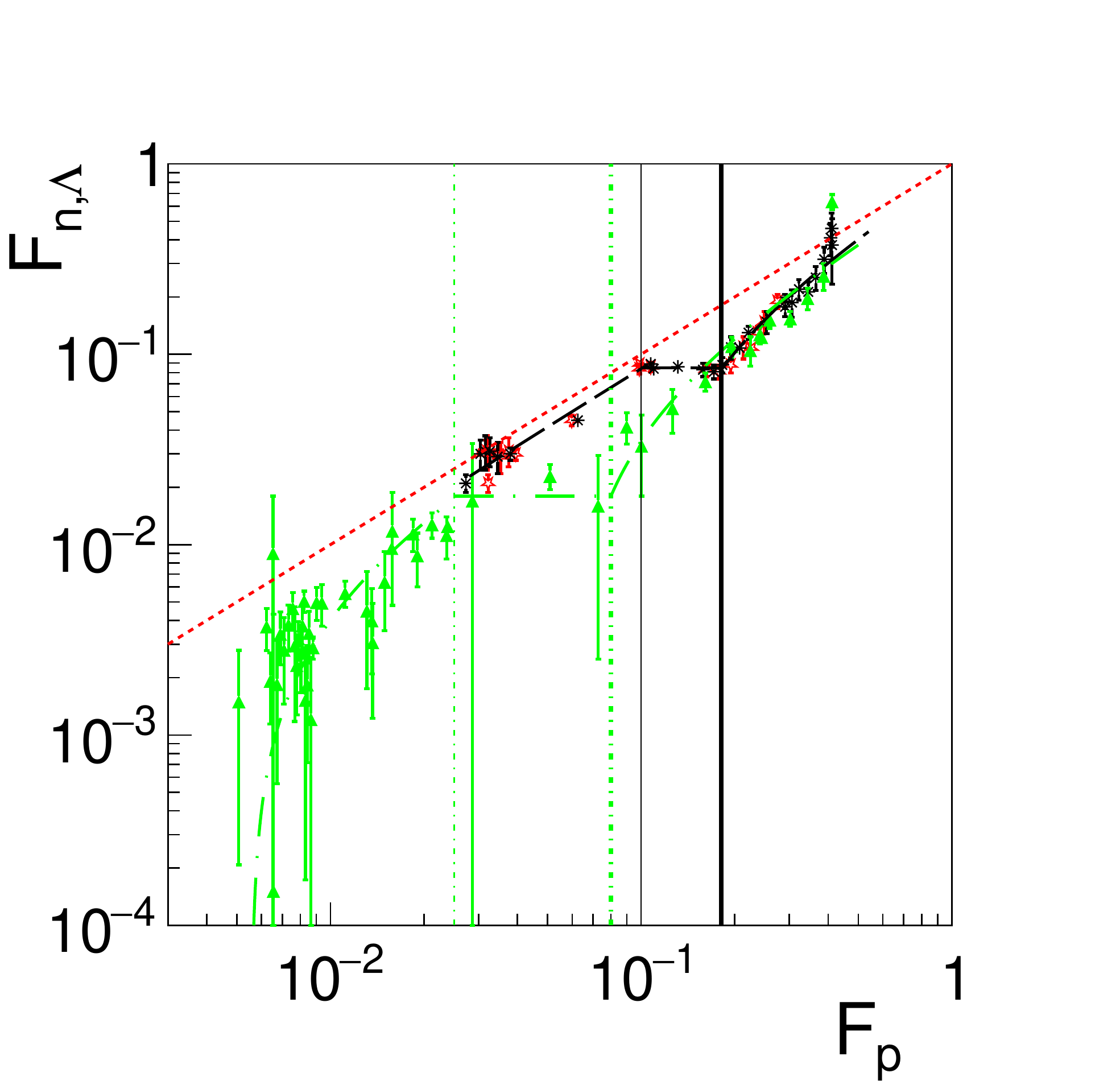} 
\caption{Neutron-proton and $\Lambda$-proton  correlation (solid green triangles). The dash-dotted thin and thick lines delimit temptatively the discontinuity region.  Other notations as in Fig. \ref{Fig:FpFn}.  } 
\label{Fig:all} 
\end{center} 
\end{figure}

Fig.  \ref{Fig:all}  shows, in addition to the proton, the  $\Lambda$ baryon data  in  logarithmic scale. 
The data for the $\Lambda$ do not have the same quality. 
Still,  they show a very similar behavior, where the change of regime 
occurs around $(F_p,F_\Lambda) \simeq (0.025, 0.018) $, i.e., 
 $p_L= 2.4$ GeV,  $(q^2=8.4$ GeV$^2$) (thin green dash-dotted line)
and the second one at 
 $( F_p,F_\Lambda,)=(0.08, 0.018)$  (thick green dash- dotted line) i.e;, 
 $p_L= 5.1$ GeV,  $(q^2=14.2 $ GeV$^2$). 
This can be associated to the production of  strange quark-antiquark pairs that are heavier and therefore correspond to a shorter time for their production and recombination.  

\section{Conclusions}
We have shown peculiar features of the baryon FF data which corroborate the picture suggested in Ref. \citep{Kuraev:2011vq} for the description of the baryon  structure. Such picture appears  coherently in scattering and annihilation reactions. 

In particular, the recent data are consistent with a neutral region at very small distances, that is responsible for a steeper decrease of the eletcric FF compared to the magnetic one. This region can be determined from the elastic scattering data to have a size smaller than 0.06 fm. If it is the case, one can predict that the FF ratio will stay small around zero. This prediction will be soon confirmed or infirmed by the planned experiments at JLab12.  

The time from the annihilation point to the hadron formation is in the range: 0.01-0.03 in units of $10^{-23}$ s,  giving a very precise inside scale  of this process. Even more precisely one can situate the transition from the  pointlike quark state to the detectable hadron,  through a complex state of different configurations.  Among them, there are overlapping configurations, with different probabilities, including diquarks, at the level of $(0.028 - 0.035) \cdot 10^{-23}$ s for nucleons and slightly shorter for strange baryons: ($0.018-0.022)\cdot 10^{-23}$s. This region corresponds to the expansion of  the quark and gluon system created from the $e^+e^-$ annihilation to  constituent quarks getting a mass and a dimension after absorbing gluons.  
A similar behavior, but with a different scale between hyperons and nucleons can be understood, as mentiond above, by the different mass of quarks involved. In the instanton picture, in Ref. \cite{Vainshtein:1982zc}, a classification of the currents and a mass scale for the formation of the different hadrons  were suggested, based on the dependence on quantum numbers of the interaction with the vacuum field.  

It is also interesting to notice that the size of the system near threshold, when, due to the competition of the between the kinetic and the confinement energies, the relative velocity of the formed hadron and antihadron leaving the interaction zone is very small,  can reach hundreds of fm. 

\section{Acknowledgments}
We acknowledge Andrea Bianconi for interesting discussions and  advices. We are grateful to Victor Kim for continuous interest in this topics.  Thanks are due to Yury Bystritskiy for useful discussions. 


\end{document}